# Post-Userist Recommender Systems
## A Manifesto

Robin Burke, Department of Information Science, University of Colorado, Boulder
Morgan Sylvester, Independent Scholar, Louisville, Colorado

**Abstract**
We define *userist recommendation* as an approach to recommender systems framed solely in terms of the relation between the user and system. *Post-userist recommendation* posits a larger field of relations in which stakeholders are embedded and distinguishes the recommendation function (which can potentially connect creators with audiences) from generative media. We argue that in the era of generative media, userist recommendation becomes indistinguishable from personalized media generation, and therefore post-userist recommendation is the only path forward for recommender systems research.

## 1. The role of creators

- In marketplaces for digital goods where recommender systems are commonly used, costs are (in the limit) driven by the portion of platform revenue shared with item providers / creators.[1]
- When digital goods are replaced by generative media, the creators' share can be eliminated and replaced by capital investment: software, servers, etc.[2]

Therefore: In a generative media setting, commercial recommender systems for digital goods are incentivized to shrink the role of creators in their ecosystems.[3]

## 2. Userist recommendation

*Userist recommendation* refers to an approach to recommender systems in which the recommendation task is solely defined in terms of the relationship between the recommender system and the recommendation consumer.[4]

---

[1] Synthetic posts for social media feeds used by Facebook and other platforms must also therefore be considered digital goods for the purposes of this discussion.

[2] Automated synthesis of apartments, restaurants, jobs, and other non-digital goods is, as yet, beyond the state of the art.

[3] A counter-argument might be that prompt engineers and others who guide generative systems to produce desirable output are the new creators. However, there is no evidence that such individuals have achieved any status as creators in most generative areas where recommender systems are applied. In general, we find that platforms tend to hide the fact that generative techniques are being used by creating fake identities for non-existent creators. This would not happen if prompt engineers were considered creative contributors.

[4] Our userist / post-userist distinction draws heavily on the work of Baumer and Brubaker: Baumer, Eric PS, and Jed R. Brubaker. "Post-userism." *Proceedings of the 2017 CHI Conference on Human Factors in Computing Systems*. 2017.

- Given the trajectory of generative AI, it is reasonable to assume that it will be possible to deliver personalized collections of synthesized content to replace current recommendation slates of digital goods.[5]
- From a userist perspective, these two options are indistinguishable: in the end, the user receives a set of items tailored to their tastes.

Therefore: Given our conclusion from Section 1, we expect that creator-free synthesized feeds will be preferred by companies operating today's recommendation platforms, and we note that there is evidence that this substitution is already starting to happen, particularly in music streaming.[6] Therefore, userist recommendation is a technology of decreasing value for digital goods in an era of generative AI.

## 3. Post-userist recommendation

*Post-userist recommendation* decenters the singular (userist) relation between recommender system and user, instead focusing on a broader set of relations. This set can include relations among users, among creators and users, between creators and the system, and among creators. Post-userist recommendation is distinguished from *multistakeholder recommendation*, which extends the set of stakeholders considered in design, evaluation, and optimization.[7] Post-userist recommendation is a relational framework, positing the importance of relations between different parties impacted by a recommender system, including but de-emphasizing the user-system relation.

- Successful post-userist recommendation designs will aim at the creation and maintenance of the creator / audience relation in addition to the system / user relation.
- The creation and maintenance of such relations is precisely the distinguishing characteristic that recommender systems can offer in a generative media era.

## 4. Conclusion

Userist recommendation has run its course as an approach to personalized access to digital goods. We are left with two options:

- Personalized information synthesis in which user profiles are used to drive the creation of media, exiling creators and yielding largely synthetic media ecosystems. We leave this dystopian topic to others.
- Post-userist recommendation, which de-centers the user/system relation and embraces the larger sociotechnical context by which media are created and consumed.

---

[5] Note that we do not consider personalized content synthesis to be a form of recommendation. These two operations have completely different semantics: retrieval from a set of pre-existing items as opposed to the creation of novel items.

[6] Hoover, Amanda. "Spotify has an AI music problem—but bots love it." Wired (2023). https://www.wired.com/story/spotify-ai-music-robot-listeners/

[7] Abdollahpouri, Himan, et al. "Multistakeholder recommendation: Survey and research directions." *User Modeling and User-Adapted Interaction* 30 (2020): 127-158.

A research agenda for post-userist recommender systems includes but is not limited to:
- Studying how recommenders support the creator / audience relation.
- Designing, implementing and evaluating creator-side interfaces to recommender systems.
- Design, implementing and evaluating community-oriented transparency and governance for recommender systems.
- Studying recommendation refusal and resistance: where does userist recommendation break down?